# Precision Spectroscopy of Nitrous Oxide Isotopocules with a Cross-Dispersed Spectrometer and Mid-Infrared Frequency Comb

D. Michelle Bailey, Gang Zhao,[†] and Adam J. Fleisher*

Material Measurement Laboratory, National Institute of Standards and Technology, Gaithersburg, MD 20899, U.S.A.

**ABSTRACT:** As a potent greenhouse gas and ozone depleting agent, nitrous oxide ($N_2O$) plays a critical role in the global climate. Effective mitigation relies on understanding global sources and sinks which can be supported through isotopic analysis. We present a cross-dispersed spectrometer, coupled with a mid-infrared frequency comb, capable of simultaneously monitoring all singly substituted, stable isotopic variants of $N_2O$. Rigorous evaluation of the instrument lineshape function and data treatment are discussed. Laboratory characterization of the spectrometer demonstrates sub-GHz spectral resolution and an average precision of $7.4 \times 10^{-6}$ for fractional isotopic abundance retrievals in 1 s.

Nitrous oxide ($N_2O$) is now the third largest greenhouse gas contributor to radiative forcing due to its strong global warming potential and prevalence in agriculture. The global average of $N_2O$ in the atmosphere has increased 20 % since the industrial revolution and emission rates from agriculture alone equate to a warming of more than 10 % of that attributed to fossil fuel production of $CO_2$.[1] To implement effective mitigation strategies, global sources and sinks of $N_2O$ must be known. Precision measurements of $N_2O$ isotopic composition are one valuable approach for determining such sources, sinks, and even mechanisms of formation.[2-10]

The singly substituted isotopocules of $N_2O$ include: $^{14}N^{14}N^{16}O$, $^{14}N^{15}N^{16}O$ ($^{15}N^{\alpha}$), $^{15}N^{14}N^{16}O$ ($^{15}N^{\beta}$), $^{14}N^{14}N^{18}O$, and $^{14}N^{14}N^{17}O$. Mass-resolving techniques were the first employed for $N_2O$ isotope retrieval.[11, 12] Initial studies focused on the site-preference of singly substituted $^{15}N$ isotopomers as a method to elucidate net sources and sinks.[2] Further correlations between total $^{15}N$ content ($^{15}N^{Bulk}$) and agricultural or natural soil sources have been reported while site preference has been critical to determine the mechanism of $N_2O$ production.[4] When used in combination with retrievals for $^{18}O$-$N_2O$, site-preference data has also been used as a tracer for $N_2O$ consumption in soils.[4]

In recent years, optical techniques have proven to be a viable alternative to less mobile mass-resolving techniques. Single frequency lasers combined with optical pathlength enhancement have been successful in determining isotopic composition of $N_2O$ in laboratory and field environments.[6-8, 10, 13-17] While technology and instrument capabilities have improved, sensitivity and spectral range can still be a limiting factor. In many cases, only a selection of isotopes can be targeted unless a second laser diode is introduced[13] and minutes of data collection may be required.[15] Implementing optical frequency combs as the light source provides a potential solution for these limitations.

Optical frequency combs are now widely used to achieve multiplexed measurements of trace gas species.[18-20] As broad bandwidth laser sources, comprising a series of discrete frequencies separated by the laser repetition rate, combs allow for high-resolution measurements over unprecedented energy and time scale. Detection schemes include dual-comb,[21] Fourier transform,[22] and spatially dispersive spectroscopies.[23] Due to the ability to collect data on microsecond timescales with no moving parts and a broad instantaneous bandwidth,[24] spatially dispersive spectroscopy was selected as the basis of the instrument presented in this work.

Comb-based spatially dispersive spectrometers have the sensitivity to enable the study of reaction kinetics critical to understanding atmospheric processes,[24, 25] but can be limited in terms of absolute accuracy when comb teeth are unresolved. Challenges in retrieving accurate optical frequency assignments and corresponding transmission through a gas sample are often attributed to a complex instrument function.[24, 26-28] In order to elevate these spectrometers for use in high-resolution spectroscopy, rigorous evaluation of the instrument function along with spectral reconstruction independent of reference databases are required. Here, we present an instrument for precision spectroscopy of $N_2O$ using an optical frequency comb centered near 4.5 μm (2222 cm$^{-1}$) coupled to a cross-dispersed spectrometer using a virtually imaged phased array (VIPA) and ruled diffraction grating. The spectrometer lineshape and optical frequency distribution has been evaluated across the focal plane array which, when incorporated into a spectral fitting routine, enables simultaneous and accurate retrieval of all singly substituted, stable isotopic variants of $N_2O$.

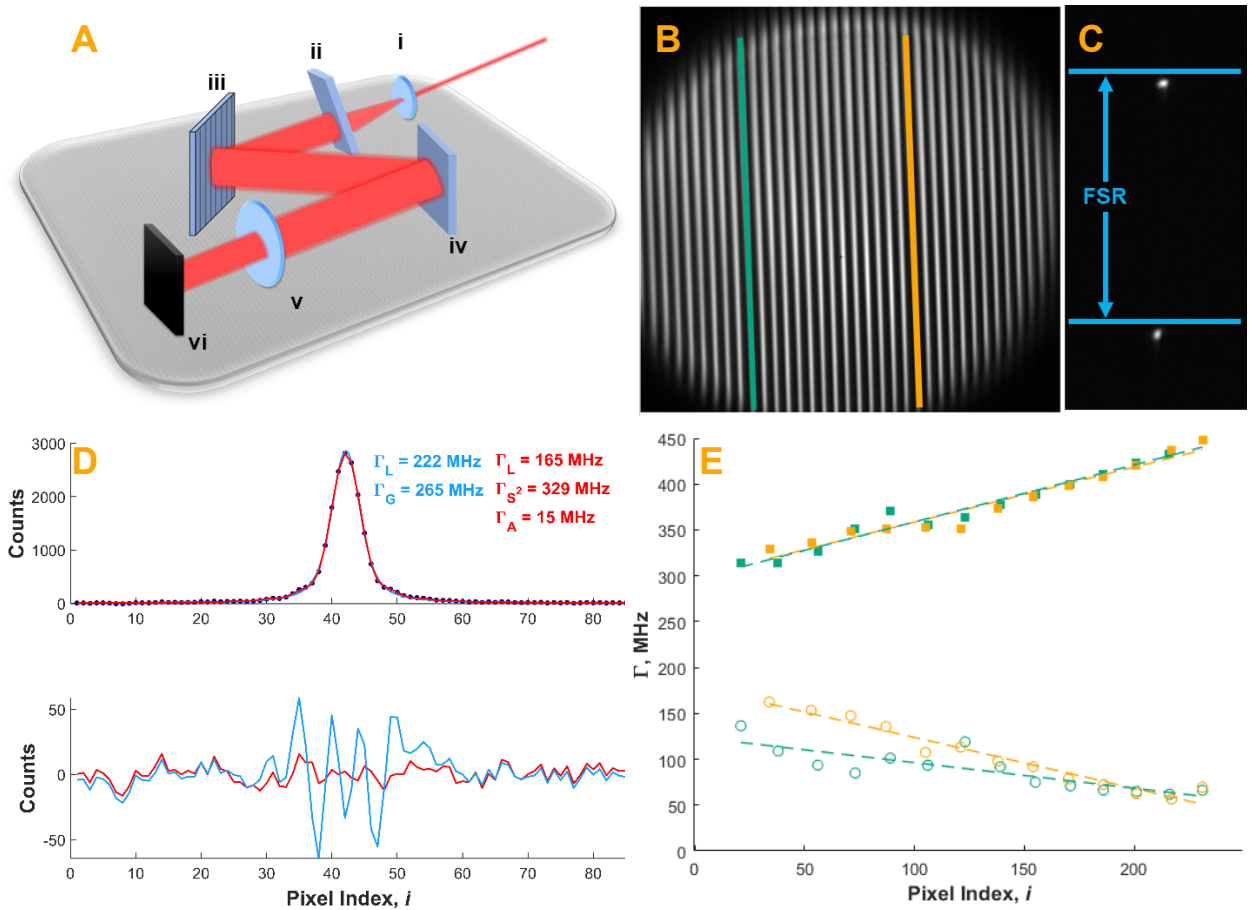

**Figure 1.** Cross-dispersed spectrometer design and characterization. A) The spectrometer optical design includes a cylindrical lens (i), VIPA etalon (ii), diffraction grating (iii), flat mirror (iv), imaging lens (v), and InSb array (vi). B) 2D image of dispersed light from the optical frequency comb. Light fringe locations used for instrument lineshape (ILS) characterization are identified with green and yellow lines. C) 2D image of dispersed light from the cw-QCL shown with approximate VIPA free-spectral-range (FSR). D) Representative fit (top) and residuals (bottom) of the cw-QCL lineshape with traditional Voigt (blue) and proposed ILS (red) models. E) cw-QCL fit results for Lorentzian (○) and Sinc$^2$ (■) contributions ($\Gamma$, half-width at half maximum) to the ILS for fringes on the left (1B, green) and right (1B, yellow) sides of the image. Not shown are contributions from the Gaussian decay, indicative of imaging asymmetry, which were nearly constant across both fringes.

## METHODS

**Spectrometer Characterization. Figure 1A** illustrates the optical design of the cross-dispersed spectrometer. Light injected into the spectrometer was collimated then line-focused with a cylindrical lens for coupling into the VIPA etalon entrance window with anti-reflective coating. The VIPA etalon dispersed light in the vertical direction while its different transmission orders were separated in the horizontal plane by a ruled diffraction grating.[29] This two-dimensional (2D) dispersion was imaged onto an InSb detector array (640 pixels by 512 pixels). Representative images for broadband comb and single-frequency laser dispersion are included in **Figure 1B-C**.

An important characteristic of dispersive spectrometers is the instrument lineshape (ILS) function imparted on the transmitted light. A rigorous understanding of the ILS is required to accurately model and subsequently evaluate transmission data from the spectrometer. Previously, the instrument lineshape imparted on molecular spectra retrieved from cross-dispersed spectrometers using VIPA etalons was assumed to be Voigt in nature.

This assumption evolved from the known Lorentzian behavior of the VIPA etalon transmission[30-32] coupled with an approximate Gaussian contribution from the grating spectrometer, particularly the pixel pitch of the focal plane array.[32] However, with ample signal-to-noise (SNR), the Voigt profile has proven insufficient for experimental results presented in this work. Here, a Lorentzian lineshape convolved with a Sinc$^2$ function was implemented along with a final convolution with a Gaussian decay (**Figure 1D**). The Sinc$^2$ function is the standard single slit function[33] often used for grating spectrometers[34] and the decay component was used to account for asymmetry in the lineshape which can be attributed to minor optical aberrations.[35] Representative curves for each ILS component and the composite function is shown in **Figure 2**. See "Spectral Model" section for further information.

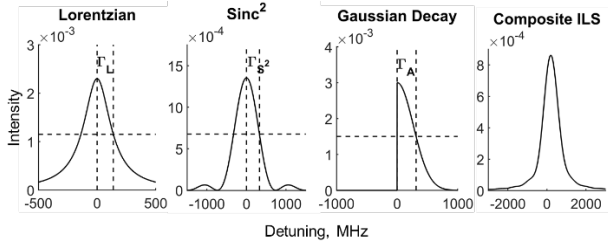

**Figure 2.** Representative functions for ILS components, illustrating defined HWHM (Γ), and the composite ILS function.

We initially evaluated the instrument lineshape by imaging a single frequency, continuous-wave quantum cascade laser (cw-QCL) at various points on the focal plane array in order to empirically determine the best-fit line profile. A total of 27 cw-QCL images, comprising two fringe locations and the full extent of the VIPA free-spectral-range (FSR) in the vertical direction, were collected. To tune the location of the cw-QCL in the vertical direction the temperature of the diode was adjusted while the diffraction grating was rotated to move the location of the incident beam in the horizontal plane. These 2D images were unwrapped as described in the following section. Each one-dimensional (1D) output was fit with the proposed lineshape and the relationship between halfwidth contributions of each function and cw-beam location was determined (**Figure 1E**). These relationships were used as initial parameters for the ILS when fitting molecular spectra.

**Spectral Acquisition.** **Figure 3** shows the free-space optical path prior to entering the spectrometer. Output from an offset-free difference-frequency generation (DFG) optical frequency comb, with a stabilized repetition rate ($f_{rep}$) of 250 MHz (commercial Rb frequency standard, short-term stability $< 2 \times 10^{-11}$ at 1 s) and full tunable bandwidth spanning 4.4 μm to 4.7 μm, propagated through a reference channel and probe channel. The probe channel held a single-pass absorption cell ($L$ = 6.985 cm) which contained the gas sample. Pressure of the sample was monitored with a digital manometer having full-scale range of 130 kPa and calibrated by a NIST secondary standard. Ambient laboratory temperature was monitored by a platinum resistance thermometer in good thermal contact with an optical mount adjacent to the sample cell. Comb light from each channel was coupled onto a single-mode fiber for injection into the spectrometer.

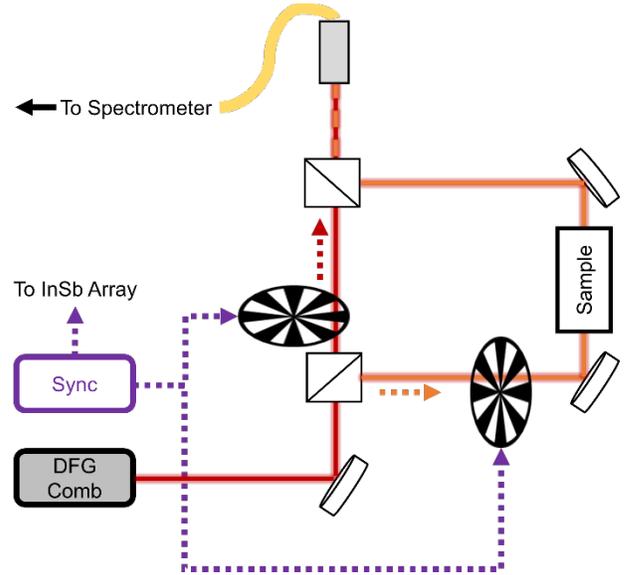

**Figure 3.** Free-space optical path. The output of the DFG comb is split to allow propagation through a reference channel and sample channel where synchronized optical choppers allow for rapid acquisition of background, reference, and probe images by the InSb array.

To enable rapid acquisition of reference and probe images, two optical choppers in the free-space path were synchronized to one-half and one-quarter of the frame rate of the InSb array (148 Hz). This allowed for sequential capture of background ($I_B$), reference ($I_R$), and probe ($I_P$) images (**Figure 4**) and retrieval of a single normalized transmission image (T) at a rate of 37 Hz with 3 ms total integration time ($t_{int}$ = 1 ms per image). To nullify image drift over the acquisition period, each image sequence was treated as follows:

$$T(i) = \frac{I_P(i) - I_B(i)}{I_R(i) - I_B(i)}, \quad (1)$$

where $I_X(i)$ is the image intensity at pixel $i$ for each image X = P, R, and B. Due to the challenge of normalizing differential spectral throughput when coupling each channel to the optical fiber, the transmission image with a target gas in the sample cell was further normalized by a transmission spectrum collected with the sample cell under vacuum.

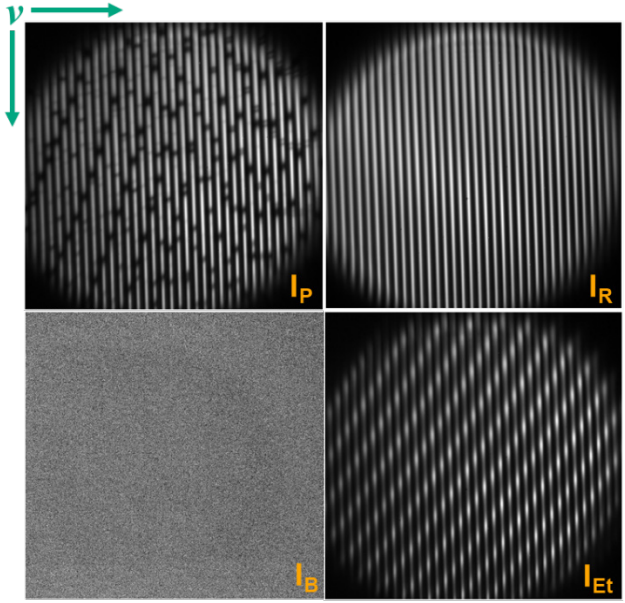

**Figure 4.** Representative images from data acquisition approach. Images shown: probe channel with sample ($I_P$), reference channel ($I_R$), background image ($I_B$), and probe channel with silicon etalon in place and no sample gas ($I_{Et}$).

Before quantitatively evaluating transmission data, the 2D image was unwrapped into a 1D spectrum illustrating optical transmission as a function of relative pixel. To achieve this, custom software was written to read in the reference channel 2D image, identify the illuminated pixel peak locations in each row of the image, and append the peak locations with corresponding transmission values in a 1D array in order of increasing frequency (as indicated by arrows in **Figure 4**). The same pixel grid was used for all subsequent probe, etalon, and cw-QCL images at that grating position. In order to compare the transmission spectrum to reference data and evaluate spectrometer performance, a method to construct an accurate frequency axis from relative pixel data was needed.

**Frequency Axis Construction.** An absolute frequency reference was combined with relative frequency markers to construct a traditional 1D transmission spectrum. The cw-QCL was coaligned with the output of the DFG comb and used to estimate the instrument lineshape function, the number of pixels in the VIPA FSR, and serve as an absolute frequency reference at a single location in InSb images (**Figure 1C**). A silicon etalon certified by NIST to have known length and FSR was placed in the probe channel optical path, with the sample cell under vacuum, to provide frequency markers throughout the 2D image (**Figure 4**) and provide a more accurate determination for the number of pixels in a single VIPA FSR.

As shown in **Figure 1C**, the VIPA FSR can be constrained in pixel space by where the image begins to repeat itself (i.e. the appearance of a second spot when imaging the single frequency cw-QCL). By unwrapping only this isolated section, a significant portion of redundant frequency information will be removed. The pixel location of the cw-QCL beam also serves as the starting point for determining optical frequency throughout the remainder of the image. Along a separate free-space optical path, the cw-QCL was mixed with the output of the DFG comb and the resulting RF beat note was monitored by a fast photodetector (>350 MHz electronic bandwidth). The frequency of the beat note was used in combination with the frequency of the cw-QCL measured by a wavemeter with manufacturer specified accuracy of ± 0.225 GHz at 300,000 GHz (or 1 μm) to determine the appropriate reference value, thus maintaining an absolute frequency grid with respect to the stabilized optical comb $f_{rep}$.

The silicon etalon with NIST-verified length (4.18929 mm ± 0.00025 mm at 296 K) and frequency-dependent FSR (10.4503 GHz at $\tilde{v} = 2222$ cm$^{-1}$ where $n = 3.4239$) was used to construct the relative frequency axis. When placed in the optical path, the 1D transmission exhibits pseudo-sinusoidal behavior where the period corresponds to the etalon FSR. The output is fit using the FSR as a constraint in order to extract the change in frequency per pixel ($dv/di$). The etalon transmission was also used to achieve accurate fringe-to-fringe stitching when transforming images to 1D data. By extending the target area slightly beyond the cw-QCL estimate, the end of one vertical etalon transmission fringe can be overlapped with the beginning of the next fringe in order to determine with sub-pixel precision the proper splicing of the two segments. This method ensured that there were no redundant frequencies or transmission data. When combined with the absolute reference point, this approach provides a frequency axis for transmission data that is independent of spectral reference databases.

**Spectral Model.** Quantitative evaluation of transmission data was achieved by combining the derived ILS characteristics with modelled molecular spectra.[18, 36] First the absorption coefficient, $\alpha(\tilde{v})$, was modelled over the target wavenumber, $\tilde{v}$, range based on the relation:

$$\alpha(\tilde{v}) = \chi \left(\frac{p}{kT}\right) Sg(\tilde{v}), \quad (2)$$

where $\chi$, $p$, and $T$ (mole fraction, pressure, and temperature) were measured, $S$ (line strength) was provided by the HITRAN2016 database, and $g(\tilde{v})$ (molecular line profile) was treated as Voigt and calculated using the same reference data.[37] The resulting $\alpha(\tilde{v})$ was sampled at $f_{rep}$ before calculating optical transmission, $T_{f_{rep}}(\tilde{v})$, defined by

$$T_{f_{rep}}(\tilde{v}) = \exp\left(-\alpha_{f_{rep}}(\tilde{v})L\right), \quad (3)$$

where $L$ is the length of the absorption cell. Finally, this calculated transmission for each comb tooth was convolved with the ILS function to yield the model $T_{ILS}(\tilde{v})$:

$$T_{ILS}(\tilde{v}) = T_{f_{rep}}(\tilde{v}) \otimes G_L(\tilde{v}) \otimes G_{S^2}(\tilde{v}) \otimes G_A(\tilde{v}), \quad (4)$$

where $G_L(\tilde{v})$, $G_{S^2}(\tilde{v})$, and $G_A(\tilde{v})$ are the area-normalized Lorentzian, Sinc$^2$, and Gaussian decay components of the composite empirical lineshape function $G_{ILS}(\tilde{v})$. The methodology is generally illustrated in **Figure 5**.

Although an expression comprising concepts described in **Eqs. (2)-(4)** appears in Cossel et al.,[18] to the best of our knowledge it has never been applied to direct frequency comb spectroscopy using a cross-dispersed spectrometer. Instead, prior works have convolved an estimated, constant ILS with the molecular line shape function $g(\tilde{v})$ in **Eq. (2)**—a substantial approximation which could lead to biases in retrieved spectroscopic parameters, e.g. mole fraction.

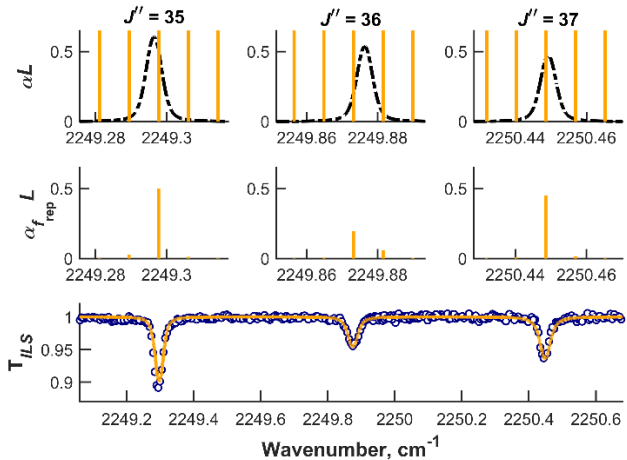

**Figure 5.** Approach for simulating molecular spectra with derived ILS. First row: HITRAN simulation of absorbance in the sample cell (dashed line, $\chi = 1$, $p = 1.35$ kPa, and $T = 297.7$ K) overlaid with a stick spectrum of unperturbed comb teeth ($f_{rep} = 250$ MHz). Second row: Simulated absorbance sampled at $f_{rep}$. Third row: Convolution of comb-sampled optical transmission with ILS (solid line) overlaid with experimental data (○).

## RESULTS AND DISCUSSION

**Model Validation.** We demonstrated the rigorous nature of our cross-dispersed spectrometer characterization and data treatment by evaluating a pure gas sample of $N_2O$. **Figure 6** shows data for the $\nu_3$ band of $N_2O$ collected over three grating positions while operating at a single crystal temperature of the DFG comb. Frequency axis construction proceeded as described above for the first two positions, where the image bandwidth encompassed the cw-QCL. The third grating position, without an absolute reference in frame, underwent the construction of a relative frequency axis as described except that spectral overlap with the preceding position was used to determine an appropriate absolute frequency. The resulting transmission spectra from each grating position were stitched together using the respective frequency axes to construct a single, full-bandwidth spectrum.

The composite spectrum exhibits rich rovibrational structure with relative intensities which deviated from the expected line-by-line Boltzmann factors, a phenomenon which is successfully mirrored in the model (**Figure 6**, inverted). We found this behavior was fundamentally linked to the $f_{rep}$ of the comb and was critically dependent on accurate knowledge of the absolute frequency and a tooth-by-tooth ILS modeling approach.

While the absolute frequency was determined through a radiofrequency (RF) beat note between the cw-QCL and the comb, the long-term stability of the free-running cw-QCL limited the measurement uncertainty to approximately 30 MHz. Therefore, the optical frequency of the cw-QCL, $\nu_{cw-QCL}$, was assigned to:

$$\nu_{cw-QCL} = nf_{rep} + f_{RF}, \quad (5)$$

where $n$ is an integer determined from the wavemeter measurement of the frequency of the cw-QCL and $f_{RF}$ is the RF beat frequency. This ensures the frequency axis assigned to experimental data is accurately linked to the stabilized comb output (i.e. integer multiples of the repetition rate). This is critical because an error smaller than $f_{rep}$, or <250 MHz, results in model spectra with distinctly different band structure due to comb teeth sampling different portions of the Doppler-broadened molecular lines, which indicates we are operating in a pseudo tooth-resolved regime. Further, if the simulations of $\alpha(\tilde{\nu})$ are not sampled at $f_{rep}$ before convolving with the ILS, the unique structure is lost altogether.

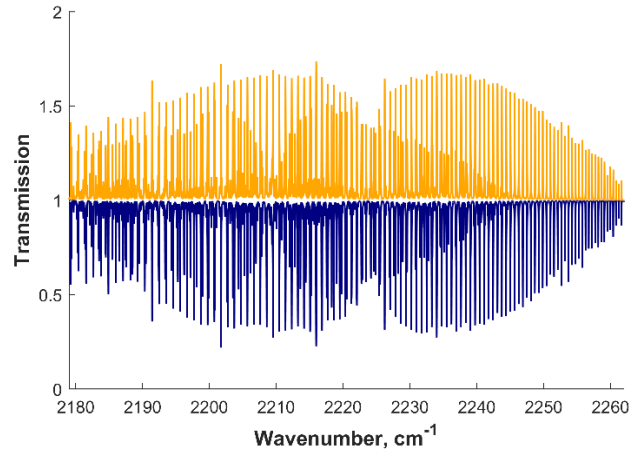

**Figure 6.** Experimental (blue) and simulated (yellow, inverted) data for pure $N_2O$ at 0.69 kPa. Simulations shown here used a single, average value for Lorentzian and $Sinc^2$ contributions to the ILS as reported in **Figure 7** and assume negligible Gaussian decay (i.e. asymmetry) contributions.

**2-D Instrument Lineshape Characterization.** Characterization of the full spectrometer ILS revealed further variability across the 2D image. Typically, with comb transmission, more than 26 vertical fringes were imaged by the InSb array in a single frame (each separated by dark pixels not exposed to incident radiation). In order to capture variation in the ILS across the image, two fringe locations were targeted with the cw-QCL (**Figure 1B**). Results in **Figure 1E** show a linear dependence of Lorentzian and $Sinc^2$ HWHM contributions to the ILS as the beam traversed the vertical fringe. Trends were consistent horizontally across the image and the HWHM of the Gaussian decay was nearly constant. Combining these results with the frequency per pixel relationship determined by etalon transmission data also allowed for ILS parameters as a function of relative frequency to be calculated.

To account for this behavior when modelling molecular spectra, a variable ILS was implemented through a fringe-by-fringe simulation approach. Transmission spectra from each grating position were split into component fringes spanning one VIPA FSR (~34 GHz). Spectra were simulated as described above on a per-fringe basis and convolved with the empirically derived ILS parameters. As the cw-QCL was not widely tunable, we were unable to evaluate the ILS at each fringe in each grating position. Therefore, an initial fit of the model to $N_2O$ data allowed the slope and intercept of the ILS relationship to float for every fringe while all $N_2O$ parameters were fixed. This resulted in a more detailed understanding of the change in each ILS component over the 2D image and is shown in **Figure 7**. A disk filter with 3-pixel radius was used to smooth the ILS data for visualization.

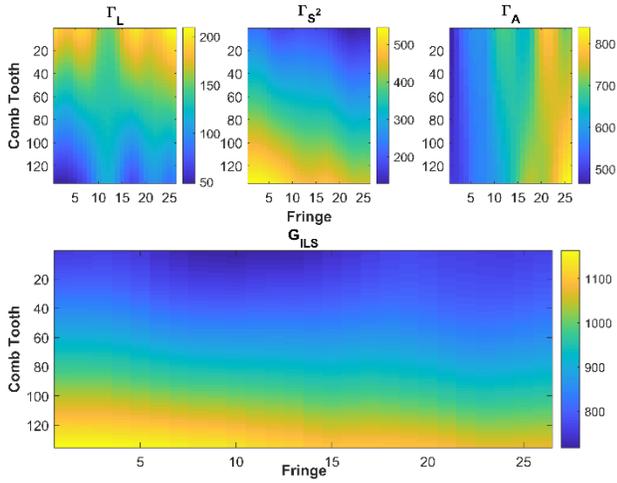

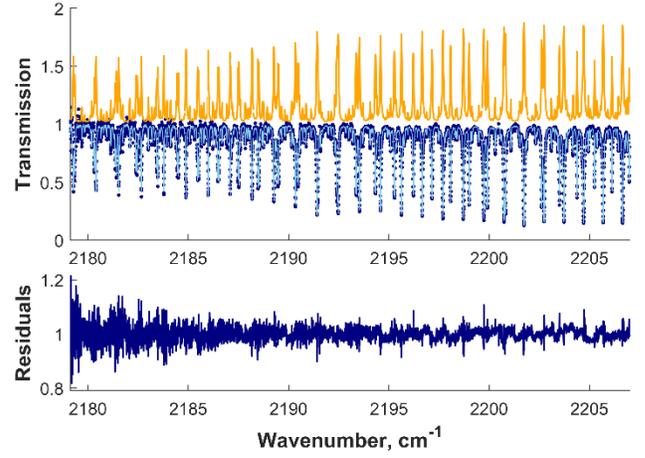

**Figure 7.** ILS characterization results after fringe-by fringe fitting with pure N$_2$O at 1.35 kPa. Top: Variation in Lorentzian, Sinc$^2$, and Gaussian decay HWHM contributions to the composite FWHM are shown for each comb tooth used in simulation over each light fringe at a single grating position. Bottom: Variation in the composite ILS FWHM.

The composite full width at half maximum (FWHM) spanned 725 MHz to 1160 MHz, slightly larger than the FWHM estimated using the VIPA etalon and cross-dispersed spectrometer design parameters (500 MHz from VIPA etalon, 350 MHz from the grating spectrometer, 680 MHz convolved total).[32] **Figure 7** shows a pronounced vertical dependence in the composite linewidth with the maximum change in FWHM ($\Delta$FWHM$_{max}$) $\leq$ 410 MHz in a single fringe while a more gradual shift occurs in the horizontal ($\Delta$FWHM$_{max}$ $\leq$ 86 MHz). Relative contributions of ILS components confirmed an anticorrelated vertical dependence for the Lorentzian and Sinc$^2$ contributions throughout the image whereas the Gaussian decay component primarily showed an increasing horizontal contribution. It is important to note that the data shown corresponds to the lowest-frequency grating position which was selected for isotopocule analysis. Here we are imaging near the edge of the comb output spectrum where the loss of photons, and thus SNR, towards the left side of the image reduces certainty in the retrieved ILS components.

**Test Case – N$_2$O Isotope Abundance Retrieval.** Highlighted in **Figure 8** are single-shot and averaged spectroscopic results. The ILS for each fringe in the spectrum was fixed to parameters shown in **Figure 7**. Then, a band-wide fit was performed to extract the fractional isotopic abundances ([N$_2$O], where [N$_2$O] can take on unitless values in the range of 0 to 1). The fitted values of [N$_2$O] were constrained such that the $\sum$[N$_2$O] over all singly substituted isotopocules was equal to the sum calculated from the HITRAN2016 fractional isotopic abundances ($\sum$[N$_2$O]$_{HT}$ = 0.99997028). **Figure 9** shows the [N$_2$O] precision, $\sigma$, from the fit.

**Figure 8.** Transmission spectrum, modelled results, and residuals for 1.35 kPa pure N$_2$O at a single grating position. Top: Inverted model (yellow), single-shot measurement (blue dots), and averaged measurement (light-blue line). Bottom: Residuals for the single-shot spectrum.

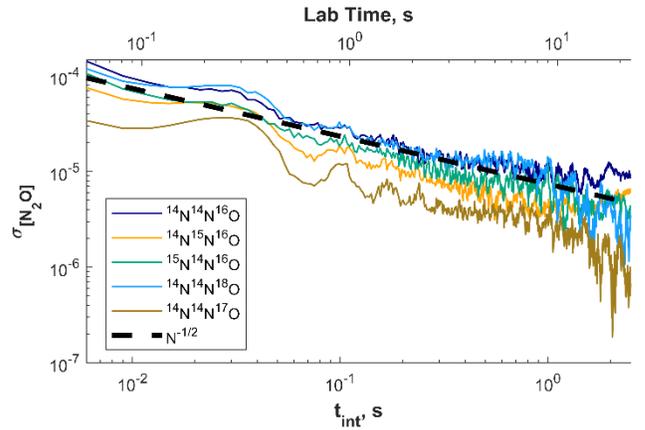

**Figure 9.** Allan deviation of fractional isotopic abundance retrievals from data shown in **Figure 8**. The dashed line indicates a representative $N^{-1/2}$ trend.

While the precision in [N$_2$O] is essential to demonstrate the utility of the broadband spectrometer, the sensitivity to changes in isotope ratios is more relevant to assess the capability of this instrument to evaluate sources and sinks of N$_2$O. Using $^{15}$N$^\alpha$ as an example, the isotope ratio, $\delta$, can be calculated as:[38]

$$\delta^{15}N^\alpha = \left( \frac{\frac{[^{14}N^{15}N^{16}O]_{exp}}{[^{14}N^{14}N^{16}O]_{exp}}}{\frac{[^{14}N^{15}N^{16}O]_{ref}}{[^{14}N^{14}N^{16}O]_{ref}}} - 1 \right), \quad (6)$$

where [N$_2$O]$_{exp}$ represents a measured fractional isotopic abundance of the sample gas and [N$_2$O]$_{ref}$ is that of a reference gas. Assuming the error in fractional abundance retrieval for each isotopocule is uncorrelated, the uncertainty in the isotope ratio ($\sigma_\delta$) can be estimated as the quadrature sum of the relative uncertainties ($\sigma_r$) in the individual [N$_2$O]$_{exp}$ and [N$_2$O]$_{ref}$ quantitates comprising **Eq. (6)**, e.g.:

$$\sigma_{r,456} = \frac{\sigma_{[^{14}N^{15}N^{16}O]}}{[^{14}N^{15}N^{16}O]} \quad (7)$$

During initial evaluation of our spectrometer no standard reference gas was available. However, we can approximate $\sigma_\delta$ using the known experimental precision if a reference gas sample of similar composition (e.g., pure $N_2O$), and therefore optical transmission, is assumed.

$$\sigma_\delta \approx \sqrt{2(\sigma_{r,456}^2 + \sigma_{r,446}^2)} \qquad (8)$$

**Table 1** lists HITRAN2016 values[37] for the fractional abundance of each isotopocule, fitted $[N_2O]_{exp}$ at maximum $t_{int}$ (10.5 s or 3,499 transmission spectra), fit precision at 1 s, and the predicted uncertainty for isotope ratio measurements at 1 s. Results are from a single grating position and data shown in **Figure 8** and **Figure 9**. Note that the relative precision achieved for the mole fraction of $^{14}N_2^{16}O$ at 1 s of lab time is ~3 × 10$^{-5}$. If a maximum frame rate of $1/t_{int}$ was achieved (i.e., zero dead-time image acquisition), using either a higher-frame-rate infrared camera or sub-windowing techniques, we could approach the 1-s figure-of-merit projected from the integration time axis of ~1 × 10$^{-5}$. Such a figure-of-merit compares favorably with the best values reported in a recent evaluation of commercial gas analyzers with significantly longer sample path length.[17]

**Table 1.** HITRAN2016 values for isotopic abundance, fitted abundance results at maximum $t_{int}$, experimental precision at 1 s, and estimated isotope ratio uncertainty at 1 s for all singly substituted isotopocules of $N_2O$.

| Isotope | $[N_2O]_{HT}$ | $[N_2O]_{exp}$ | $\sigma_{[N_2O],1s}$, $10^{-5}$ | $\sigma_\delta$, $10^{-3}$ |
|---|---|---|---|---|
| $^{14}N^{14}N^{16}O$ | 0.99033300 | 0.990117 | 1.1 | n.a. |
| $^{14}N^{15}N^{16}O$ | 0.00364100 | 0.003728 | 0.58 | 2.2 |
| $^{15}N^{14}N^{16}O$ | 0.00364100 | 0.003758 | 0.82 | 3.1 |
| $^{14}N^{14}N^{18}O$ | 0.00198600 | 0.002091 | 0.92 | 6.2 |
| $^{14}N^{14}N^{17}O$ | 0.00036928 | 0.000276 | 0.26 | 13 |

CONCLUSIONS

We have demonstrated the utility of a cross-dispersed spectrometer coupled with a mid-infrared frequency comb for simultaneous retrieval of all singly substituted, stable isotopic variants of $N_2O$. Presented is a detailed characterization of the instrument lineshape function and a method for robust frequency axis construction to achieve precision spectral modelling. We validated our approach for frequency axis construction by observing comb-tooth sampling of Doppler-broadened rovibrational lines without the use of molecular reference data as frequency markers. Comb-tooth by comb-tooth implementation of the variable ILS reduced our fit residuals by more than an order of magnitude—further validating spectrometer linearity and frequency accuracy. Precision fractional isotopic abundance retrievals indicate that the spectrometer provides adequate sensitivity with 1 s integration time and would be useful in evaluating pure $N_2O$ samples from various sources;[4, 10] however, sensitivity would need to improve via an increase in the sample path length in order to report real-time variation.[6, 8]

Future work will focus on developing a more robust scheme for instrument lineshape characterization. The ILS is directly affected by changes in optical alignment, some of which are induced by temperature change altering the dispersive properties of the VIPA etalon. As a result, a rapid evaluation of the ILS with minimal impact to the operational status of the spectrometer is required. While it is possible to employ a filter cavity in the free-space optical path to spatially resolve comb modes on the InSb array,[23] that approach would lead to a substantial loss of signal reaching the spectrometer and ultimately degraded data quality. Instead, a Fabry-Pérot QCL operating in a harmonic state with GHz spacing[39] could provide frequency markers and a multi-point reference grid for the ILS. Integrating a device of this kind would enhance the data throughput and enable future autonomous operation of the spectrometer.

Designing highly dispersive mid-infrared optics to fully resolve comb teeth at standard repetitions rates of 80 MHz to 250 MHz represents an ideal alternative. Although substantial losses in the mid-infrared coatings required to create high-finesse VIPA etalons have previously hindered research in this direction, Roberts et al.[40] have recently reported a cross-dispersed VIPA spectrometer with record resolution of 190 MHz in the C-H stretch region from 3.0 μm to 3.5 μm — almost a factor of 5 better than the resolution we report at the center of the spectrometer image (890 MHz) for a wavelength of 4.5 μm. However, whenever the spectrometer resolution is of the same order of magnitude as $f_{rep}$, the ILS may result in measurable crosstalk between imaged comb teeth, and therefore should still be thoroughly considered over the entire image. Here we required a comb-tooth-resolved model for accurate spectral analysis of our Doppler-broadened absorption lines, even with a center-image resolution of $3.6 f_{rep}$. Therefore, robust approaches for rigorous ILS analysis remain important for rapid and accurate molecular spectroscopy using optical frequency combs and 2-D cross-dispersed spectrometers.

AUTHOR INFORMATION

**Corresponding Author**

*E-mail: adam.fleisher@nist.gov

**Permanent address**

†Institute of Laser Spectroscopy, State Key Laboratory of Quantum Optics and Quantum Devices, Shanxi University, Taiyuan City 030006, Shanxi Provence, P. R. China

ACKNOWLEDGMENTS

The authors would like to thank Dennis Everett of the Dimensional Metrology Group at NIST for providing precision measurement of the silicon etalon length. Dr. Erin Adkins (NIST) and Dr. Kevin Cossel (NIST) commented on the manuscript.